\documentstyle[12pt]{article}

\date{}
\begin{document}

\title{Diffeomorphism Invariance of Geometric Descriptions of Palatini and 
Ashtekar
Gravity \thanks{%
this work is supported by NSF of China, Pan Den Plan of China and LWTZ -1298
of Chinese Academy of Sciences}}
\author{{Yan Luo\thanks{{\protect\small E-mail: Luoy@itp.ac.cn}}, Ming-Xue 
Shao%
\thanks{{\protect\small E-mail: shaomx@itp.ac.cn}}, Zhong-Yuan Zhu\thanks{%
{\protect\small E-mail: zzy@itp.ac.cn}}} \\
{\small Institute of Theoretical Physics, Academia Sinica, Beijing 100080,
P.R.China.}}
\maketitle
\date{}
\begin{abstract}
In this paper, we explicitly prove the presymplectic forms of the Palatini
and Ashtekar gravity to be zero along gauge orbits of the Lorentz and
diffeomorphism groups, which ensures the diffeomorphism invariance of these
theories.\\

PACS: 04.20.Cv; 04.20.Fy \\Keywords: Diffeomorphism invariance; presymplectic
forms; Palatini; Ashtekar
\end{abstract}

\vskip 0.6in

Geometric description and quantization \cite{Wood Sniatycki} is the global
generalization of ordinary Hamiltonian canonical description and
quantization. This formulism has been shown to provide an natural way to
investigate global and geometrical properties of physical systems with
geometrical invariance, such as Chern-Simons theory \cite{Witten-cs}, anyon
system \cite{YZL}, and so on. But the traditional descriptions of
geometrical and canonical formalism of classical theories are not manifestly
covariant, because from the beginning one has to explicitly choose a
``time'' coordinate to define the canonical conjugate momenta and the
initial data of systems. Several year's ago, E.Witten \cite{Witten} and
G.Zuckerman \cite{Zuckman} and C.Crnkovic \cite{CE} et al. suggested a
manifestly covariant geometric description, where they took the space of
solutions of the classical equations as phase space. This definition is
independent of any special time choice so that is manifestly covariant. Then
they used this description to discuss Yang-Mills theory, string theory and
general relativity etc. Recently, B.P.Dolan and K.P. Haugh \cite{DH} used
Crnkovic and Witten's method to deal with the Ashtekar's gravity. They
investigated the problems related to the complex nature. But a thorough
discussion needs to prove the vanishing of components of presymplectic form $%
\omega $ tangent to the diffeomorphism and Lorentz group orbits, as Crnkovic
and Witten emphasized in ref. \cite{CE} for the case of general relativity.
Essentially they pointed this proof is the most delicate point in their
treatment. Therefore this short paper is devoted to complete this proof for
cases of Palatini and Ashtekar gravity.

The first order action of Palatini with the tetrads and Lorentz connections
as its configuration space variables is given in \cite{Ashtekar-book}: 
\begin{equation}
S_P(e,\omega )=\frac 12\int_{}^{}R_{ab}^{IJ}e_I^ae_J^bed^4x,
\end{equation}
where $e_I^a$ 's are tetrads and $e$ is the determinant of $e_a^I$. The
curvature of the Lorentz connections $\omega _a^{IJ}$ is defined to be $%
R_{ab}^{IJ}=\partial _a\omega _b^{IJ}-\partial _b\omega _a^{IJ}+[\omega
_a,\omega _b]^{IJ}$. Here $^{``}a,b,c,d ,...^{"}$ stand for Riemannian
indices and ``$I,J,K,L,...$'' stand for the internal $SO(3,1)$ indices. The
variations of the action with respect to $\omega _a^{IJ}$ and $e_I^a$ give
the equations of motion: 
\begin{equation}
e_I^cR_{cb}^{IJ}-\frac 12R_{cd}^{MN}e_M^ce_N^de_b^J=0,
\end{equation}
\begin{equation}
\partial _be_I^a+\omega _{bI}^{~~J}e_J^a+\Gamma _{bc}^ae_I^c=0,
\end{equation}
in which $\Gamma _{bc}^a$ is the Christoffel. The second equation of motion
(3) can be written in the form: 
\begin{equation}
\bigtriangledown _be_I^a=0,
\end{equation}
in which $\bigtriangledown $ is torsion-free connection on both space-time
and internal indices. By using equation (4), the first equation of motion
(2) becomes 
\begin{equation}
R_{ab}=0,
\end{equation}
which is just the Einstein field equation in vacuum.\\The tangent vectors of
the solution space satisfy the linearized equations of motions, 
\begin{equation}
2\bigtriangledown _{[c}\delta \omega _{b]}^{IJ}e_I^c=-R_{cb}^{IK}\delta
(e_I^ce_{aK})e^{aJ},
\end{equation}
\begin{equation}
\bigtriangledown _b\delta e_I^a+\delta \omega _{bI}^{~~ J}e_J^a+\delta
\Gamma _{bc}^ae_I^c=0.
\end{equation}
>From the action (1), we get the presymplectic form 
\begin{equation}
\Omega =\int_\Sigma \delta \omega _b^{IJ}\wedge \delta (e_I^ae_J^be)d\Sigma
_a,  \label{*3}
\end{equation}
where $\Sigma $ is the space-like supersurface in space-time manifold.
Obviously the presymplectic form $\Omega $ is independent of the choice of
the space-like supersurface $\Sigma$ and invariant under Riemannian
coordinate and Lorentz transformations. But as pointed in \cite{CE}, we need
to prove the degeneracy along gauge orbits of Lorentz and diffeomorphism
groups and, in fact, the proof is not trivial.

Under infinitesimal Lorentz transformations, the tetrads and the connections
transform as 
\[
e_K^c\mapsto e_K^c-\alpha _K^{~~J}e_J^c , 
\]

\[
\omega _b^{IJ}\mapsto \omega _b^{IJ}+\bigtriangledown _b\alpha ^{IJ},
\]
where $\alpha $ is valued in local Lorentz Lie algebra which means $\alpha $
will vanish at infinity, so we obtain the transformations of $\delta e_K^c$
and $\delta \omega _b^{IJ}$ along the Lorentz group orbits, 
\begin{equation}
\delta e_K^c\mapsto \delta ^{^{\prime }}e_K^c=\delta e_K^c-\alpha
_K^{~~J}e_J^c,  \label{*2}
\end{equation}
\[
\delta \omega _b^{IJ}\mapsto \delta ^{^{\prime }}\omega _b^{IJ}=\delta
\omega _b^{IJ}+\bigtriangledown _b\alpha ^{IJ}.
\]

>From  (\ref{*2}) and (\ref{*3}), and keeping only the terms up to first
order of $\alpha $, 
\begin{equation}
\Omega^{^{\prime }}-\Omega=\bigtriangleup \Omega =\int_\Sigma 
\bigtriangledown _b\{\alpha ^{IJ}\wedge
[\delta (e_I^ae_J^b)-e_I^ae_J^be^K_c\delta e_K^c]\}ed\Sigma _a.  \label{*1}
\end{equation}
where we have used equations (4), (7) and the antisymmetry of $IJ$ in $%
\alpha ^{IJ}$. Clearly the integrand of right hand side of equation (\ref{*1}%
) is of the form $\bigtriangledown _b(X^{ab})e$ with $X^{ab},$ an
antisymmetric tensor which makes the integral to be a surface integral at
infinity. Since we restrict the local transformations in limited region, the
surface integral vanishes identically. So, the presymplectic form (\ref{*3})
is degenerate along Lorentz group orbits.

Next, we concentrate on the proof of the degeneracy of $\Omega $ along
diffeomorphism directions. The diffeomorphism transformations are in the
form 
\[
x^a \mapsto y^a =x^a +\xi ^a ,
\]
\[
e_K^c(x)\mapsto e_K^b(y)\frac{\partial x^c}{\partial y^b}.
\]
\\ From the second equation of motion (3) and noticing $\bigtriangledown
_b\xi ^c=\partial _b\xi ^c+\Gamma^{c}_{bd}\xi^d$, one obtains 
\[
e_K^c\mapsto e_K^c-\bigtriangledown _b\xi ^ce_k^b-\omega _{d
K}^{~~~J}e_J^c\xi ^d 
\]
and similarly,  
\[
\omega _b^{IJ}\mapsto \omega _b^{IJ}+\bigtriangledown _d \omega
_b^{IJ}\xi ^d +\omega _d ^{IJ}\bigtriangledown _b\xi ^d -[\omega
_d ,\omega _b]^{IJ}\xi ^d ,
\]
so that 
\begin{equation}
\delta e_K^c\mapsto \delta ^{^{\prime }}e_K^c=\delta e_K^c-\bigtriangledown
_b\xi ^ce_k^b-\omega _{d K}^{~~~J}e_J^c\xi ^d   \label{*5}
\end{equation}
\[
\delta \omega _b^{IJ}\mapsto \delta ^{^{\prime }}\omega _b^{IJ}=\delta
\omega _b^{IJ}+\bigtriangledown _d \omega _b^{IJ}\xi ^d +\omega _d
^{IJ}\bigtriangledown _b\xi ^d -[\omega _d ,\omega_{b}]^{IJ}\xi ^d .
\]

The presymplectic $\Omega $ can be rewritten in the form 
\[
\Omega =\int_\Sigma j^aed\Sigma _a, 
\]
\begin{equation}
j^a=\delta \omega _b^{IJ}\wedge [-e_I^ae_J^be_c^K\delta e_K^c+\delta
e_I^ae_J^b+e_I^a\delta e_J^b].  \label{***}
\end{equation}
>From (\ref{*5}) and (\ref{***}) and keeping only the terms up
to first order of $\xi $, after tedious calculations, we obtain 
\[
j^{^\prime a}-j^a=\bigtriangleup j^a=\bigtriangleup j_1^a+\bigtriangleup 
j_2^a+\bigtriangleup
j_3^a, 
\]

\[
\bigtriangleup
j^{a}_{1}=\bigtriangledown_{d}[-\delta\omega_{b}^{IJ}e^{d}_{J}e^{a}_{I}%
\wedge\xi^{b}+\delta\omega_{b}^{IJ}e^{b}_{J}\wedge(e^{a}_{I}%
\xi^{d}-e^{d}_{I}\xi^{a})-\omega_{b}^{IJ}e^{d}_{J}e^{a}_{I}\xi^{b}e^{K}_{c}%
\wedge\delta e^{c}_{K}+ \omega_{b}^{IJ}\xi^{b}\wedge(e^{a}_{I}\delta
e^{d}_{J}-e^{d}_{I}\delta e^{a}_{J})], 
\]

\[
\bigtriangleup j^{a}_{2}=R_{bd}^{IJ}[\frac{1}{2}(e^{b}_{J}\delta
e^{d}_{I}-e^{d}_{I}e^{bM}e_{cJ}\delta
e^{c}_{M})\wedge\xi^{a}+e^{d}_{I}e^{aM}e_{cJ}\delta e^{c}_{M}\wedge\xi^{b}
-e^{a}_{I}e^{b}_{J}e^{K}_{c}\delta e^{c}_{K}\wedge\xi^{d}+ \delta
e^{a}_{I}e^{b}_{J}\wedge\xi^{d}], 
\]

\begin{equation}
\bigtriangleup j_3^a=-\delta \Gamma _{bc}^a\wedge e_J^ce_L^b\omega
_d^{JL}\xi ^d,
\end{equation}
where we have used equations (3) (6) (7) and the antisymmetry of {\sc IJ} in 
$\omega ^{IJ}.$ From the deformation of the equations of motion $%
e_I^cR_{cb}^{IJ}=0$, we have $\bigtriangleup j_2^a=0$. Obviously due to the
symmetry of $b$ $c$ in $\delta \Gamma _{bc}^a$ and the antisymmetry of {\sc %
JL} in $\omega _d^{~JL}$, $\bigtriangleup j_3^a=0$. So, there only leaves
with 
\[
\bigtriangleup \Omega =\int_\Sigma \bigtriangleup j_1^aed\Sigma _a. 
\]
\\Like the integrand of right hand side of (\ref{*1}), $\bigtriangleup j_1^a$
is again of the form $\bigtriangledown _dX^{da}$ with $X^{da}$ being an
antisymmetric tensor, one gets 
\begin{equation}
\bigtriangleup \Omega =\int_\Sigma \partial _d(eX^{da})d\Sigma
_a=\int_{\partial \Sigma }eX^{da}dS_{da}.
\end{equation}

If assuming $\xi ^\lambda $ has compact support or, more generally, is
asymptotic at infinity to a killing vector field (a more detailed discussion
on boundary conditions can be found in \cite{ABR}), $\bigtriangleup \Omega $
obviously vanishes which ends our proof of the degeneracy of the
presymplectic form (8) of Palatini gravity along the directions of
diffeomorphism transformations. If denote $Z$ the solution space of
equations of motion, G$_1$ the diffeomorphisms group and $G_2$ the Lorentz
group, the presymplectic form (8) is a well defined symplectic form on the
moduli space $Z/G_1/G_2$ which means that the system has constraints
corresponding to diffeomorphisms transformation and local Lorentz
transformation.

The same procedures of proof is suitable for Ashtekar gravity \cite{Ashtekar}
\cite{Ashtekar-book}. Since the only difference is that in Ashtekar's case
tetrads and connections are complex and self-dual (or anti-self-dual) which
does not change the proof, so that we arrive at the conclusion that the
diffeomorphism invariance of above geometrical description is also correct
for Ashtekar gravity.

\end{document}